\def\year{2020}\relax
\documentclass[letterpaper]{article} 
\usepackage{aaai20}  
\usepackage{times}  
\usepackage{helvet} 
\usepackage{courier}  
\usepackage[hyphens]{url}  
\usepackage{graphicx} 
\usepackage{verbatim}
\usepackage{mathptmx}
\usepackage{amsfonts}       
\usepackage{amsmath} 
\usepackage{bm} 

\title{Learning Occupational Task-Shares Dynamics for the Future of Work}

\author{Subhro Das\textsuperscript{\rm 1}, Sebastian Steffen\textsuperscript{\rm 2}, Wyatt Clarke\textsuperscript{\rm 1}, Prabhat Reddy\textsuperscript{\rm 1}, \\ 
{ \Large \bf Erik Brynjolfsson\textsuperscript{\rm 2}, Martin Fleming\textsuperscript{\rm 1} } \thanks{This work was supported by the MIT-IBM Watson AI Lab. Corresponding authors: Subhro Das (subhro.das@ibm.com), and, Martin Fleming (fleming1@us.ibm.com).}\\ 
\textsuperscript{\rm 1}MIT-IBM Watson AI Lab, IBM Research \\ 
\textsuperscript{\rm 2}Massachusetts Institute of Technology \\
}

\begin{document}

\maketitle

\begin{abstract}
The recent wave of AI and automation has been argued to differ from previous General Purpose Technologies (GPTs), in that it may lead to rapid change in occupations' underlying task requirements and persistent technological unemployment. In this paper, we apply a novel methodology of dynamic task shares to a large dataset of online job postings to explore how exactly occupational task demands have changed over the past decade of AI innovation, especially across high, mid and low wage occupations. Notably, big data and AI have risen significantly among high wage occupations since 2012 and 2016, respectively. We built an ARIMA model to predict future occupational task demands and showcase several relevant examples in Healthcare, Administration, and IT. Such task demands predictions across occupations will play a pivotal role in retraining the workforce of the future.
\end{abstract}

\section{Introduction}
\label{sec:intro}
\begin{figure*}[h!]
    \centering
    \includegraphics[width=0.85\linewidth]{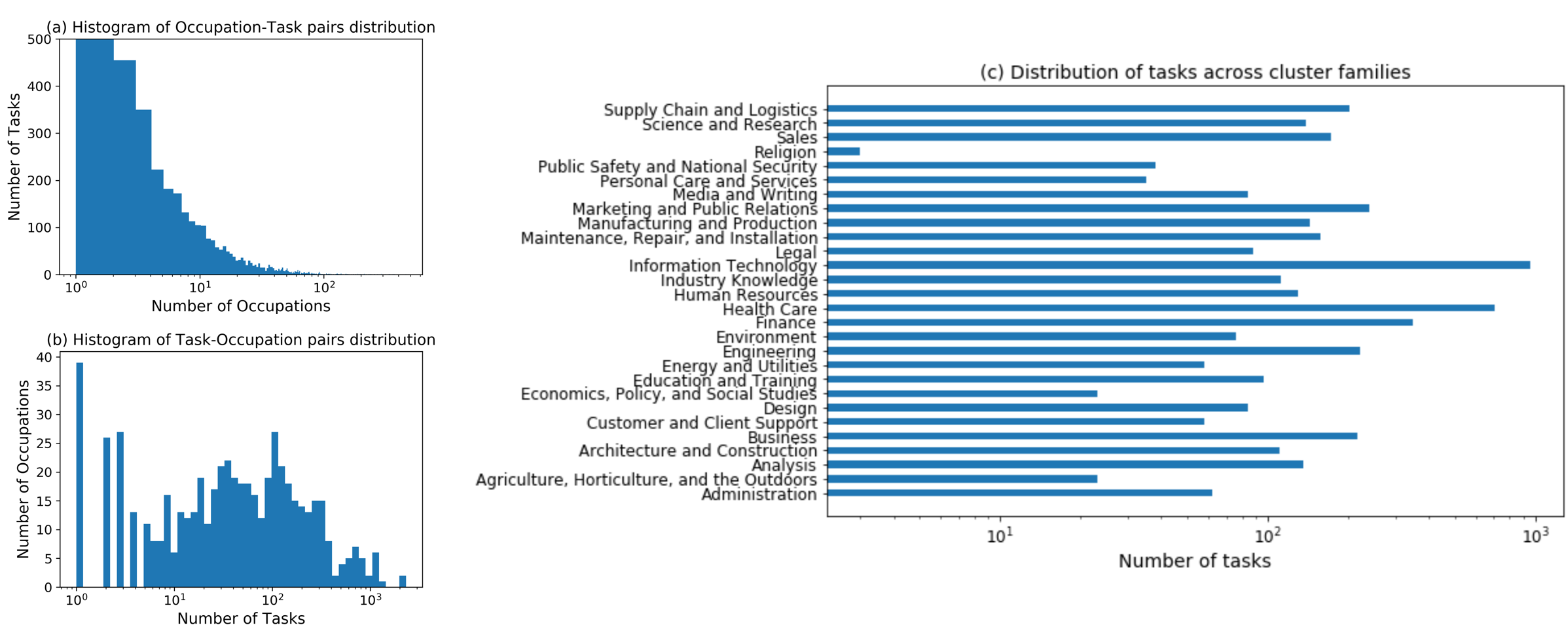}
    \caption{Histograms demonstrating summary statistics for Occupation-Task pairs data distributions.}
    \label{fig:Skill2SOC}
\end{figure*}
Artificial Intelligence, and automation more generally, is widely believed to be the next big General Purpose Technology (GPT) \cite{Brynjolfsson2018b}. Thus, it has the capacity to transform entire economies, societies, and workers' lives and occupations. Specifically, automation has the ability to: (i) make labor more productive (labor-augmenting automation), (ii) make automation itself ever more productive (automation at the intensive margin), (iii) introduce new tasks into the economy, or (iv) displace a wide range of human tasks (automation at the extensive margin) \cite{Acemoglu2019}. It has been suggested that this race between man and machine may lead to a rise of technological unemployment if automation outpaces the creation of new tasks and new occupations \cite{Acemoglu2018c}. Conversely, slow automation may not raise economic output enough and may thus not be an optimal growth path either. But no matter whether automation or (task) innovation `wins' \footnote{in parallel to the race between education and technology \cite{Goldin2007}.}, both forces lead to changes in occupations' underlying task requirements. This paper studies how occupations' specific task demands have changed over the last decade by leveraging a large dataset of online job postings. Using a novel methodology we document trends in occupations and tasks as well as occupational wage terciles (low, medium, high). 
\par
In fact, some of these changes have already manifested themselves. Some argue that the terms routine and non-routine characterize the relationship between tasks/skills and information technology (IT) and find that occupations have shifted towards requiring more analytical and interactive tasks and away from requiring cognitive-routine and manual-routine tasks \cite{Spitz-Oener2006}, especially during the period of 1950-2000 \cite{atalay2019evolution}. Skills, as a form of task-specific human capital, are an important source of individual wage growth \cite{Gathmann2010}. Thus, the relative loss of productivity of routine skills translates to lower wages and an overall more polarized wage and employment share distribution \cite{AutorDorn2013}. For several occupations, in particular low-wage ones, AI is predicted to outperform humans within the next decade  leading to significant risks of long-term unemployment \cite{grace2018will} \cite{depredicting}.
\par
And yet, adoption of automation technologies and corresponding tasks may be slow. It took almost thirty years before the design of factories changed from being centered around one GPT, the steam engine, to the single-story layout we know today that optimizes for another GPT, electricity \cite{BrynjolfssonMcAfee2014}. Some authors claim that the current wave of automation is different \cite{duckworth2019inferring}.\footnote{See \cite{wajcman2017automation} for an accessible overview.}
In particular, low wage workers may suffer the brunt of the occupational changes, productivity and wage losses as well as layoffs, since their occupations consist of a larger share of routine tasks. This Routine-Biased Technological Change (RBTC) implies that recent technological change is biased toward replacing labor in routine tasks \cite{GoosManningSalomons2014}. 

However, medium- and high-wage occupations are not immune to occupational change either. Occupations that heavily rely on IT tasks have been shown to change faster due to rapid software innovation \cite{Hershbein2018}. These fast obsoletion rates of specific software tasks lead to relatively flatter earnings profiles for STEM workers \cite{Deming2018a}. Some have argued for a `great reversal' in demand for cognitive task and shown that more educated workers have begun to crowd out less educated workers, due to sorting and changes in relative productivity of workers and capital \cite{BeaudryGreenSand2015}. Automation and IT capital, such as Data-Driven Decision Making (DDD), have been rapidly adopted and have made plants more productive and efficient, requiring even managers and other high-wage occupations to adapt to stay productive \cite{Brynjolfsson2016}, \cite{Bartel2007}. These results suggest that retraining is both necessary as well as costly, in particular for low-wage workers, and that the evolution of occupational task demands are an important phenomenon to predict and study \cite{atalay2019evolution}. With the advent of new AI technologies \cite{hemamou2019hirenet} that predicts the hirability of the candidates as evaluated by recruiters based on salient socials signals, the future job candidates needs to be better prepared to demonstrate their ability to execute the required tasks.
\par
In this paper we document recent trends in task demands across multiple dimensions, including occupations and wages by leveraging a novel large data set of online job postings between 2010 and 2017. We also predict how the demands and wages for different tasks evolve over time.

\section{Occupation and Tasks}
\label{sec:data}
All occupations can be viewed as bundles of a multitude of tasks performed by workers in that occupation \cite{Acemoglu2018c}. On the demand side, the employers define the tasks that needs to be executed by an employee in the job. Whereas, on the supply side of the labor market, the employees come with skills, the capabilities to carry out the required tasks in the job. In an occupation, the workers receive wages based on the skills that they bring in. However, when engaged in an occupation, the workers are required to perform a number of tasks. The wage earned, then, is the weighted average of the wage paid for performing a collection of tasks and providing a portfolio of skills. This distinction between tasks and skills is important when tasks can be accomplished by workers with a range of skill levels, workers in differing locations, or substituting capital for labor. In this paper, tasks will be considered to study how occupations are transforming. 

\subsection{Data}
\label{subsec:data} Our data comes from Burning Glass Technologies (BGT)\footnote{https://www.burning-glass.com/}, an analytics software company that provides real-time data on job growth, skills demands, and labor market trends. The data covers about 170 million online job vacancy postings posted on over 40,000 distinct online job sites in the United States between 2010 and 2017 and arguably covers the near-universe of job postings. Each vacancy posting is parsed and annotated with the posting date, the Standard Occupational Classification (SOC) code\footnote{https://www.bls.gov/soc/}, and which tasks were demanded, among several other variables. The tasks data is parsed via BGT's industry-leading taxonomy, which covers around $17,000$ tasks, which are nested within 572 task clusters and 28 task cluster families. For example, Python is a task within the Scripting Languages task cluster, which itself falls into the Information Technology task cluster family. This data is ideal for these purposes because it encodes jobs as bundles of tasks \cite{Deming2018}. 

There is some ambiguity as to whether the content of job postings describe skills of workers or tasks workers are required to perform. Because firms do not know workers' skills before hiring - ex ante - and because firms know with near certainty the tasks workers are to perform, in what follows the requirements will be referred to as tasks.\footnote{Job postings do not always reflect workers’ roles precisely. Especially in tight labor markets, the eventual responsibilities of workers might differ from intentions at hiring. In additions, postings can also reflect marginal rather than average occupational changes. The marginal changes can reflect replacement demand as well as net new demand.} Such a distinction is consistent with the theory that tasks are specified by employers on the demand side and skills are the capabilities workers bring on the supply side.\footnote{Because there are differences between the taxonomies, Burning Glass has not merged their skills taxonomy with the O*NET taxonomy of tasks. Some tasks in the O*NET taxonomy are not mentioned in Burning Glass postings, as they are assumptive of the position to be filled. Also, the O*NET technology tasks are not updated frequently while the Burning Glass data is updated monthly.}

\subsection{Task-Occupation Pairs}
\label{subsec:task-occ}
The Burning Glass job postings data can be represented in three-dimensions: occupations, tasks, and years. Each job posting is mapped to one of the 964 unique occupations, as defined by 6-digit Standard Occupational Classification (SOC) code by the Bureau of Labor Statistics (BLS) of the U.S. Department of Labor. The tasks, required to be performed by a worker as mentioned in the posting, are extracted and tied to the mapped occupation (SOC) for that posting. This method attributed to enumerate the number of times a task has been mentioned for a particular occupation within a given period of time. The summary statistics of this frequency data for task-occupations pairs are in Fig.~\ref{fig:Skill2SOC}.

Fig.~\ref{fig:Skill2SOC}(a), shows a histogram of task appearances across occupations (SOC). The minimum and maximum number of occupation that a task has been associated to are $1$ and $460$, respectively. There are $15$ tasks that are mentioned in more than $300$ occupations, as listed in Table~\ref{table:task2soc}, whereas, there are $3,976$ tasks that occur in fewer than 10 occupations. Some of the tasks that appear in only one occupation are:  {\it Plastic Industry Knowledge}, {\it Polymer Synthesis}, {\it Polish}, {\it Aromatherapy}, {\it Poetry}, {\it E-Procurement}, {\it Planters}, {\it Physician Sales}, {\it Plant Biology}, {\it Pizza Delivery}, {\it Aircraft Electrical Systems}, {\it Piping Replacement}, {\it Hbase}, {\it Airframe \& Powerplant}, {\it Construction Documentation}, etc.

Fig.~\ref{fig:Skill2SOC}(b) shows a histogram for the opposite mapping, i.e. the number of occupations associated with binned tasks. The minimum and maximum number of tasks that associated to an occupation are $1$ and $2312$, respectively. There are nine occupations that have more than 1,000 unique tasks mentioned in their job postings, see Table~\ref{table:soc2task}. Seven out these nine occupations are in the {\it Management}, and, {\it Computer \& Mathematical} occupation families, with occupations `Software Developers, Applications' (SOC: 15-1132) and `Managers, All Other' (SOC: 11-9199) reporting even more than $2,000$ tasks. On the other end, there are $148$ occupations which requires less than $10$ unique tasks, with $39$ among those asked for only one unique task in their postings. Most of these jobs are in the {\it Transportation \& Material Moving}, {\it Production}, {\it Construction \& Extraction}, and, {\it Installation, Maintenance, \& Repair} occupation families. This could be due to the fact that there weren't many posting related to these occupations in our data or those jobs actually require one task.

The $964$ unique occupations, represented by 6-digit occupation codes, can be categorized into $22$ occupation families represented by the first 2-digits of their 6-digit SOC codes, see \cite{fleming2019future} for details. There are $539$ unique task cluster family and occupation family pairs. Fig.~\ref{fig:Skill2SOC}(c) shows the number of unique tasks that belongs to each of the $28$ Task Cluster Families.
\begin{figure*}[t]
\centering
  \includegraphics[width=0.95\linewidth]{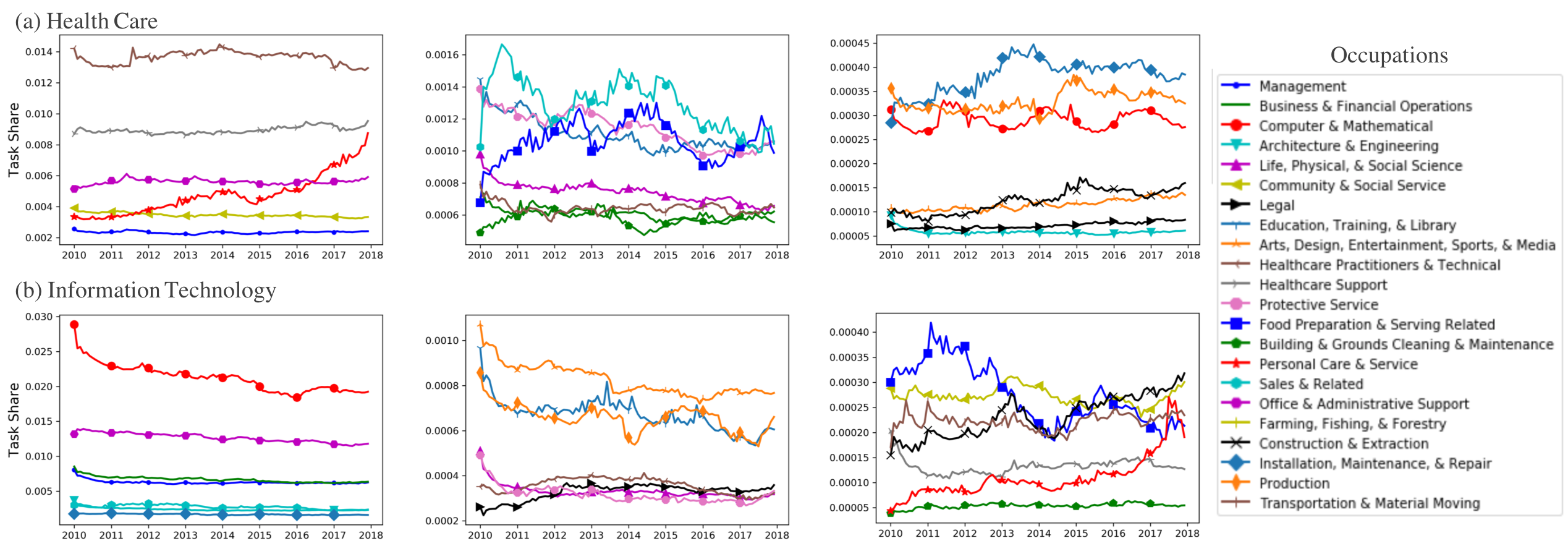}
  \caption{Task-Share dynamics of (a) Healthcare, \&, (b) Information Technology task cluster family across occupation families.}
  \label{fig:SCF_health_IT}
\end{figure*}
%

\section{Methodology: Task-Share Dynamics}
\label{sec:methods}
To understand how the occupations are evolving, we dive deeper into how tasks within them are changing. From the job postings, we get the occurrence frequency of each task in a given occupation. Using the tasks count in postings for each occupation, a time-series dataset is generated. This measures the demand from employers for workers who can perform these tasks. We incorporate wages and employment shares data from the Bureau of Labor Statistics (BLS), who publish annual statistics of the average wages and number of employees in each of the 964 occupations. We normalize the task demand time-series data by the share of workers employed in that occupation to derive the unique task-shares dynamics data for each task-occupation pair. The changes in the occupations during that period are characterized via the evolution of the task-shares within each occupation.

\subsection{Monthly Task-Share Time-Series}
\label{subsec:monthly_taskshare}
Let's denote a task by $x_i$, where $x_i \in \mathcal{X} = \{ x_1, \hdots, x_i, \hdots, x_{|\mathcal{X}|}\}$, and, $|\mathcal{X}|$ is the total number of unique tasks in the economy. An occupation is denoted by $o_j$, where $o_j \in \mathcal{O} = \{ o_1, \hdots, o_j, \hdots, o_{|\mathcal{O}|}\}$, and, $|\mathcal{O}|$ is the total number of unique occupations. Let, $t$ denote the monthly time index from January 2010 to December 2017, i.e., $t \in \mathcal{T} = \{ 01\text{-}2010, \hdots, 12\text{-}2017 \}, with |\mathcal{T}|=96 $. With these notations, the count of mentions of task~$x_i$ in occupation~$o_j$ in month~$t$ is represented by $n_{i,j,t} \in \mathbb{Z}^{+}$. Similarly, let $m_{j,t} \in \mathbb{Z}^{+}$ denote the count of mentions of occupation~$o_j$ in month~$t$. 

Under the assumption that the distribution of tasks demanded in a job listing reflects the distribution of tasks performed by workers in the corresponding occupation, we calculate the share of workers in each occupation that perform each task. The occupation-task share, $z_{i,j,t} \in \mathbb{R}^{+}$, is:
\begin{align}
    \label{eqn:z}
     z_{i,j,t} = \frac{n_{i,j,t}}{m_{j,t}}, \qquad \forall i, j, t.
\end{align}
To normalize the occupation-task share with an external baseline, we use the annual statistics of the average hourly wage and number of employees in the 964 SOC occupations published by the BLS. A piece-wise linear interpolation function was employed for converting the annual statistics to monthly statistics in order to obtain hourly wages, $w_{j,t} \in \mathbb{R}^{+}$, and number of employees, $E_{j,t} \in \mathbb{Z}^{+}$, for each occupation~$o_j$ month~$t$ combination. The share of the labor force,~$e_{j,t} \in \mathbb{R}^{+}$, employed in each occupation in the U.S. can be calculated by,
\begin{align}
    \label{eqn:e}
    e_{j,t} = \frac{E_{j,t}}{\sum_{j} E_{j,t}}, \qquad \forall j, t.
\end{align}
While online job postings account for a significant share of recruiting activity during 2010-2017, their share is increasing over time. Moreover, job listings may be biased towards white-collar jobs and may not perfectly represent current employer demands, such that these data are not necessarily representative of the US labor force. Hence, we combine the BLS employment share~$e_{j,t}$ with the Burning Glass occupation-task share~$z_{i,j,t}$ to compute the overall share of workers performing task~$x_i$ as part of occupation~$o_j$ in month~$t$ as,
\begin{align}
    \label{eqn:y}
     \bm{ y_{i,j,t} } = e_{j,t} \times z_{i,j,t}, \qquad \forall i, j, t.
\end{align}
For the rest of this paper, we will refer to this occupation-task employment share~$y_{i,j,t} \in \mathbb{R}^{+}$ as \emph{task-share} -- a time-varying metric for each task~$x_i$ performed in an occupation~$o_j$. Using this metric, we created an unique time-series dataset containing \emph{task-share}~$y_{i,j,t}$ of all the tasks across all occupations over the period of 96 months from January 2010 to December 2017. To the best of the authors’ knowledge, this is a first-of-its-kind dataset that presents the task-shares at a monthly frequency for each task-occupation pair. 
\begin{figure*}[t!]
  \includegraphics[width=0.9\linewidth]{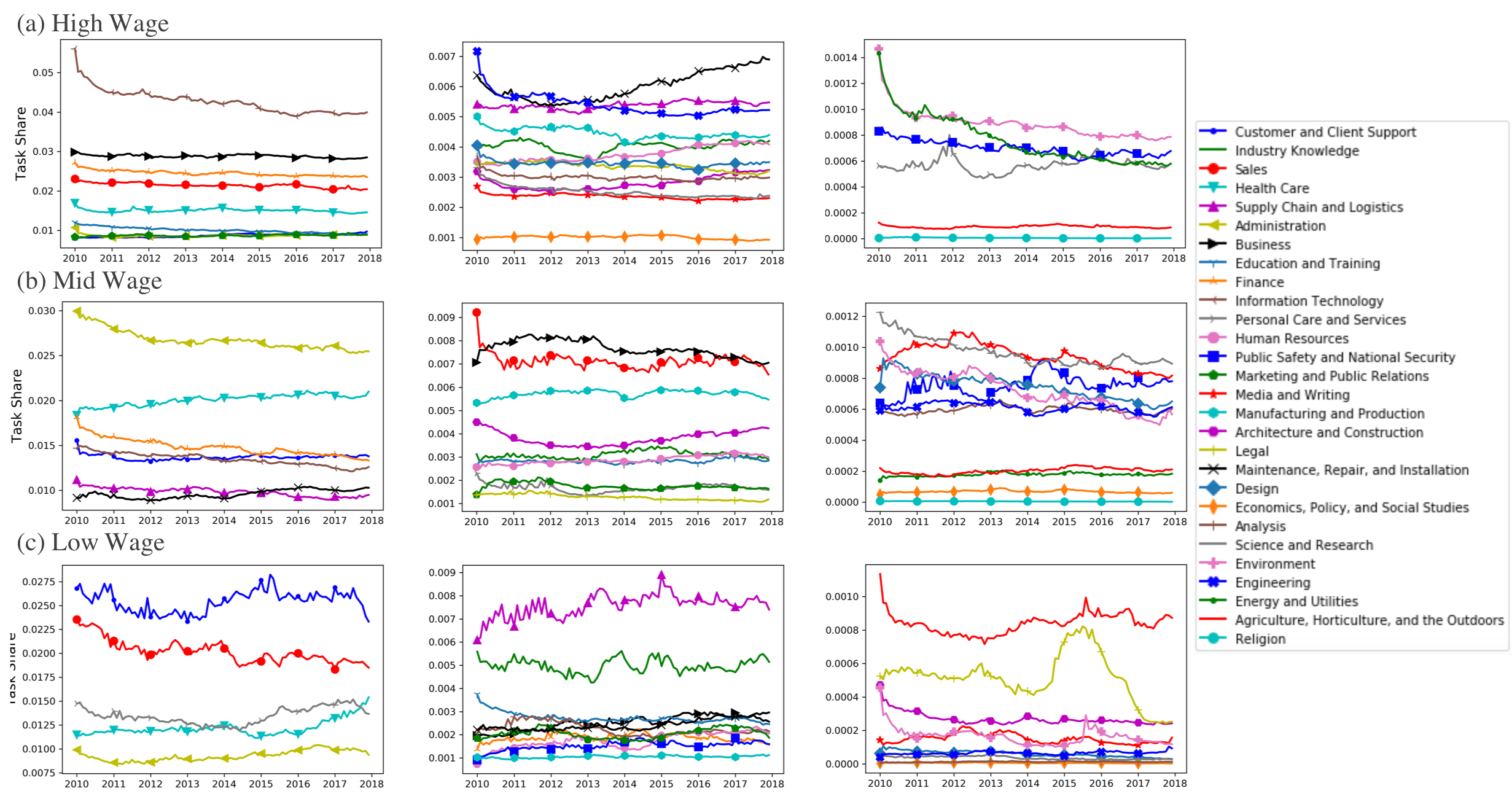}
  \caption{Task-Share dynamics of task cluster families across: (a) high, (b) medium, and, (c) low wage occupation groups.}
  \label{fig:SCF_HML}
\end{figure*}
\subsection{Task-Share Aggregation}
\label{subsec:aggregate_taskshare}
For further analyses and to extract insights on how the task-share dynamics are impacting the evolution of the occupations in the U.S. labor market, this large time-series dataset on task-occupations pairs needs to be aggregated.  We aggregate the task-shares of all task-occupation pairs at a task cluster family and occupation family levels denoted by~$\overline{y}_{p,q,t}$. Let, $\bar{x}_p$ denote a task cluster family, where $\bar{x}_p \in \mathcal{\bar{X}} = \{ \bar{x}_1, \bar{x}_2, \hdots, \bar{x}_p, \hdots, \bar{x}_{|\mathcal{\bar{X}}|}\}$ and $|\mathcal{\bar{X}}|=28$ is the total number of unique task cluster families. Similarly, an occupation family is denoted by $\overline{o}_q$, where $\overline{o}_q \in \mathcal{\overline{O}} = \{ \overline{o}_1, \overline{o}_2, \hdots, \overline{o}_q, \hdots, \overline{o}_{|\mathcal{\overline{O}}|}\}$ and $|\mathcal{\overline{O}}|=22$ is the total number of occupation families. Then, the aggregated task-share~$\overline{y}_{p,q,t}$ of workers performing tasks from task cluster family~$\overline{x}_p$ as part of occupations from occupation family~$\overline{o}_q$ in the month~$t$ is,
\begin{align}
    \label{eqn:ybar}
     \bm{ \bar{y}_{p,q,t} } = \sum_{ i,j \; : \; x_i \in \overline{x}_p, o_j \in \overline{o}_q } y_{i,j,t}, \qquad \forall p, q, t.
\end{align}
This aggregated task-share~$\overline{y}_{p,q,t}$ helps to visualize and interpret how the demand for a particular family of tasks have evolved across different occupation families, or, how the task-shares of different cluster families of tasks have evolved within a particular occupation family.

We further aggregate the task-shares data among the high, mid, and low (HML) wage occupation terciles, denoted by~$\tilde{y}_{p,r,t}$, to understand how the task-shares of different task cluster families have evolved across wage-based occupation groups. Using the average of the BLS hourly wage~$w_{j,t}$ from year 2010, the 964 occupations~$o_j$ are categorized into three wage bins, $\tilde{o}_r \in \{ \text{low, mid, high} \}$. Thus the task-share~$\tilde{y}_{p,r,t}$ of workers performing tasks from task cluster family~$\overline{x}_p$ as part of occupations from occupation tercile~$\tilde{o}_r$ in month~$t$ is,
\begin{align}
    \label{eqn:ytilde}
     \bm{ \tilde{y}_{p,r,t} } = \sum_{ i,j \; : \; x_i \in \overline{x}_p, o_j \in \tilde{o}_r } y_{i,j,t}, \qquad \forall p, r, t.
\end{align}
The downstream analyses results using these task-shares, $y_{i,j,t}$, as well as the aggregated task-shares, $\overline{y}_{p,q,t}$ and $\tilde{y}_{p,r,t}$, are presented in the following section.

\section{Results and Discussions}
\label{sec:results}
The impact of technology on labor markets has long been an important issue for economic theory, empirics, and policy. Perhaps even more important to those that make up the labor market employers and employees is that the advent of Artificial Intelligence (AI) will shift the demand for labor skills. It is imperative to understand the extent and nature of the changes so that we can prepare today for the jobs of tomorrow. While most jobs will change as AI and new technologies continue to scale across businesses and industries, so far we mainly see task shifts \textit{within} occupations instead of their disappearance. In this study, we focus on how occupations are transforming by studying the evolution dynamics of the task-shares that compose the jobs.

\subsection{Task Reorganization among Workers}
\label{subsec:res_task_reorg}
Among the 28 task cluster families, we show in Fig.~\ref{fig:SCF_health_IT} how aggregated \emph{task-shares}~$\overline{y}_{p,q,t}$ of two example cluster families, $x_p = $ \emph{Health Care} and \emph{Information Technology}, have evolved between 2010-2017 across different (2-digit SOC) occupation families. To remove noise, to leverage finely-grained variation between time steps and to better expose the task-shares, we employed a moving average smoothing function with a window of 3 months for all the task-share figures. The growth and decline rates of the task-shares is measured in terms of normalized coefficients by fitting a linear regression to the task-shares \cite{bishop2006pattern}. 

The health care task cluster family has its highest shares in `Healthcare Practitioners \& Technical', `Healthcare Support', `Office \& Administrative Support', `Personal Care \& Service', and, `Community \& Social Service' occupations (in order of demand). On the other end, its lowest shares are in `Architecture \& Engineering', `Legal', `Construction \& Extraction', and `Arts, Design, Entertainment, Sports, \& Media' Occupations. These findings are in line with what one would expect and are easily extendable to other cases. Based on the regression coefficients in Table~\ref{table:SCF_healthcare_IT_coeff}, it is evident that the healthcare task-share has seen a significant growth in `Personal Care \& Service' occupation, along with considerable growths in `Legal', `Construction \& Extraction', and `Arts, Design, Entertainment, Sports, \& Media' occupations and decline in `Sales \& Related' jobs.

In Fig.~\ref{fig:SCF_health_IT}(b), the Information Technology (IT) task cluster family has its highest shares in `Computer \& Mathematical Operations', `Office \& Administrative Support', `Business \& Financial Operations', and `Management' occupations, with declining demand in `Computer \& Mathematical Operations' occupations. IT has its lowest, yet steadily-growing shares in `Personal Care \& Service' and `Construction \& Extraction' occupations, as in Table~\ref{table:SCF_healthcare_IT_coeff}. These results are consistent with the anecdotal evidence of increased IT penetration of a variety of occupations as well as IT being a GPT. 

\subsection{High and Low Wage Jobs are Gaining Tasks}
\label{subsec:mid_wage}
In the interest of studying how task-shares of different task cluster families have evolved across occupations with different wages levels, in Fig.~\ref{fig:SCF_HML} we display the evolution of aggregated \emph{task-shares}~$\tilde{y}_{p,q,t}$ across wage terciles (low, medium, high). The top five task-shares for high wage occupations are `Information Technology', `Business', `Finance', `Sales', and `Health Care'; for mid-wage occupations they are `Administration', `Health Care', `Finance', `Customer \& Client Support', and `Information Technology'; and for low-wage jobs they are `Customer \& Client Support', `Sales', `Personal Care \& Services', `Health Care', and `Administration'. Although the `Maintenance, Repair, \& Installation' and `Human Resources' task cluster families had small task-shares in both high-wage and low-wage occupations, they still saw a steady and significant growth in demand. Comparable growth also happened for `Architecture \& Construction' \& `Customer \& Client Support' in high-wage jobs, and, `Business' \& `Public Safety \& National Security', `Engineering' in low-wage jobs. The regression coefficients in Table~\ref{table:SCF_HML_coeff} provide additional details. Notably, for mid wage occupations, most task cluster families experienced declines. Such a transition in the task-shares among wage-based occupation groups indicates that mid wage occupations are losing shares overall, and that task-shares in high and low wage occupations are growing. This evidence of a more polarized workforce is consistent with the U-shaped occupational share and wage patterns found in Autor, Dorn (2013).

\subsection{AI and Related IT Technologies}
\label{subsec:res_IT}
\begin{figure}[t!]
  \includegraphics[width=0.95\linewidth]{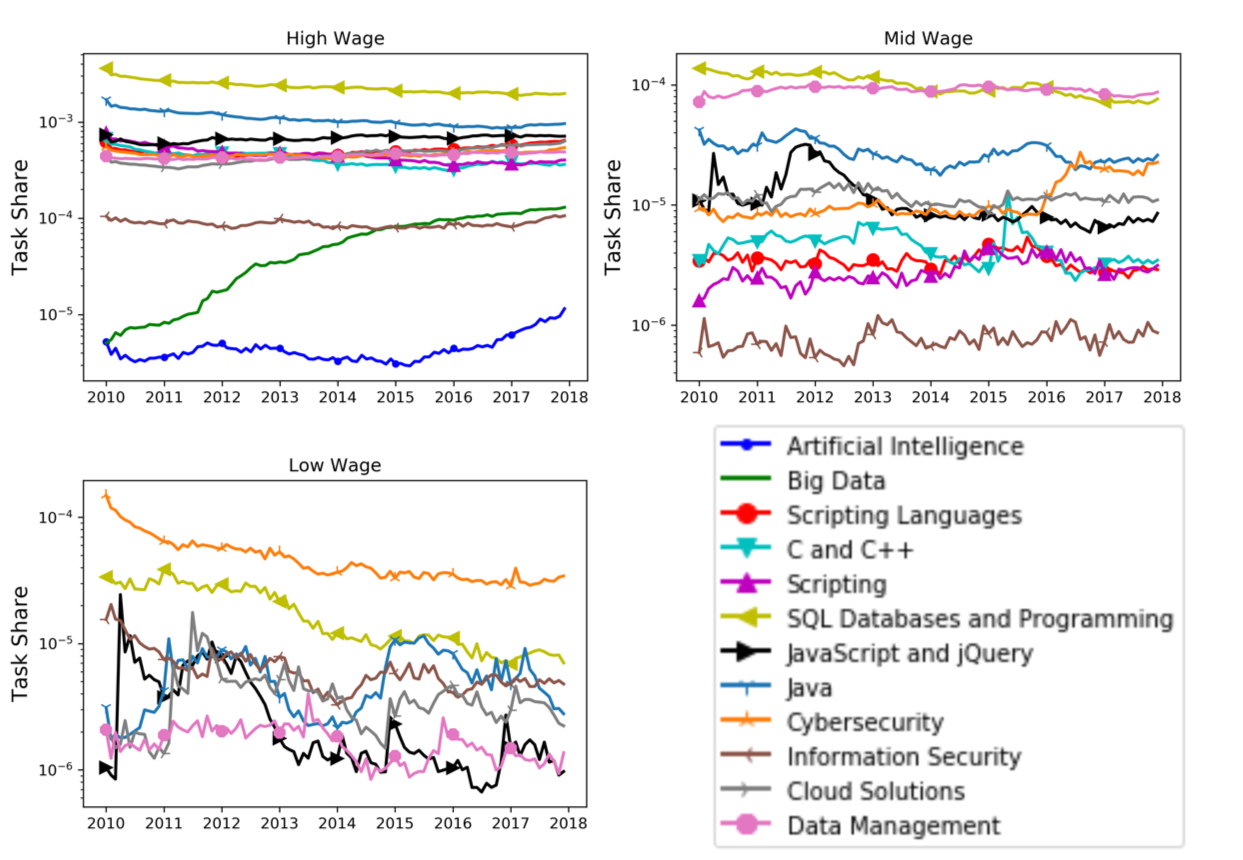}
  \caption{Task share dynamics of different Information Technology task clusters across HML wage occupations.}
  \label{fig:SC_IT}
\end{figure}

To study how AI and related technologies are impacting the labor market at the initial phase of adoption, we zoom into the Information Technology (IT) task cluster family to look at specific task clusters. In Fig.~\ref{fig:SC_IT}, we plot the task-shares of selected task clusters within the IT task cluster family across high, mid and low (HML) wage occupations. Although the `SQL Databases and Programming', `Java' and `JavaScript \& jQuery' task clusters have the highest shares in high and mid wage occupations, their demand is steadily declining, see  Table~\ref{table:IT_HML_coeff}. In contrast, even though the `Artificial Intelligence' and `Big Data' task clusters had low task-shares in the high wage occupations, their demand increased at a very high rate during 2010-2017. These task-cluster have not seen any demand in the mid and low wage occupations. On the one hand, task clusters like `Scripting Languages' (includes Python) and `Cloud Solutions' are gaining task-shares in high wage occupations. On the other hand, most IT task clusters are losing task-shares in low wage occupations. This evolution of IT task demands confirms the industry trends towards developing AI-based products and services in the Cloud requiring workers to perform AI, Big Data, Scripting Languages, and Cloud Solutions based tasks while focusing less on traditional software products and services that require workers to perform SQL, Java, and Data Management oriented tasks.

\subsection{Task-Share Forecasting}
\label{subsec:res_forecast}
In addition to the insights already extracted, this study and dataset lays down the scope and foundation for detailed exploration of the evolution of occupations (and the tasks within) across different industries in the US labor market. The task-shares time-series data creates an opportunity to learn the dynamics of task and occupations, and, then quantitatively predict the task-shares for near future with confidence bounds. Such predictive capabilities on the labor market might help the workers reskill themselves, corporations retrain their employees, or, new graduates to learn the skills to be able to execute the tasks of the future.

\begin{figure}[t!]
  \includegraphics[width=0.96\linewidth]{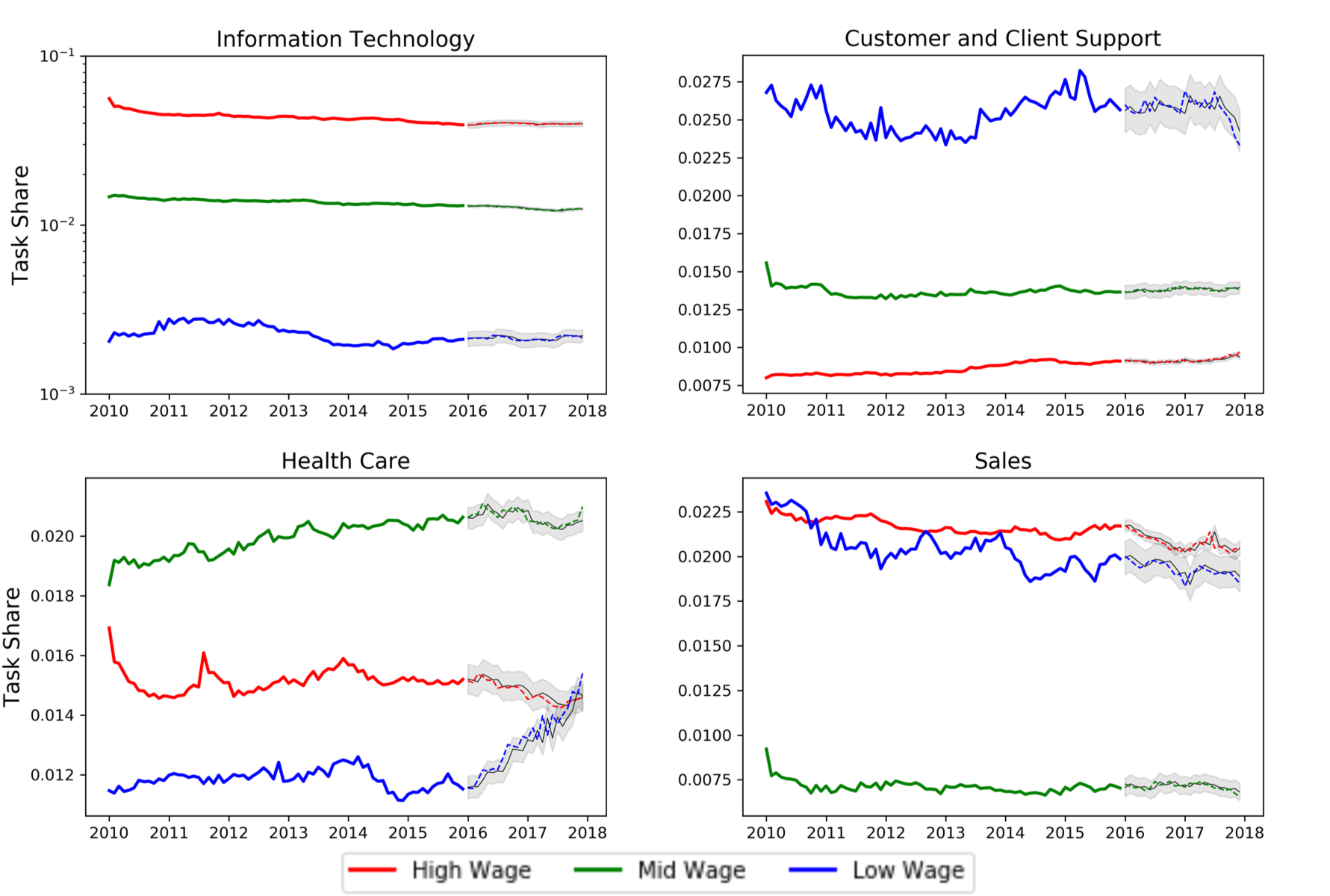}
  \caption{One-step ahead predictions of task-shares of selected task clusters families across HML wage occupations.}
  \label{fig:pred}
\end{figure}

In the first phase of this study, we have trained an autoregressive integrated moving average (ARIMA) model~\cite{makridakis2008forecasting} to learn the representation dynamics of the task-shares of different task cluster families across HML wage occupations over the first 72 months of data (2010-2016). Using this trained ARIMA model, we make one-month ahead predictions of the task-shares. The mean absolute percentage error (MAPE) of predictions is considerably less than~$5\%$ in most cases as shown in Table~\ref{table:TS_HML_pred}. In Fig.~\ref{fig:pred}, we plot the task-share forecasts (black lines) with $95\%$ confidence intervals (grey areas) to compare against the true task-shares (dotted lines) for a few selected task cluster families across high (red line), mid (green line), and low (blue line) wage occupations. The accuracy of the task-share predictions is a clear indicator towards the benefit of developing robust and more accurate forecasting models to characterize the evolution of occupations and the tasks therein. 
%

\section{Conclusions \& Next Steps}
Some of the task trends are striking. Notably, the fast rise of Big Data and Artificial Intelligence in high wage occupations since 2012 and 2016, respectively. This delayed, yet rapid development seems similar to the adoption of electricity in the 1890s as well computers in the 1970s - both started slow and labor productivity growth did not take off for over twenty years \cite{BrynjolfssonMcAfee2014}. Thus, we may have another decade or so giving workers ample time to adapt with the occupational transformation.


\par
This empirical research sheds new light on the transformation of work by characterizing occupations in terms of task-shares dynamics. There are still many open questions remaining in the study. To extract further empirical evidence as to what is occurring in the US labor market, it would be crucial to investigate: (a) how task-share dynamics are evolving across different industries and across different geographical/Metropolitan regions within the country; (b) dynamic functional coupling between different task-shares across occupation groups; and, (c) impact of task-share dynamics on wage-dynamics and vice versa. Today, we know the change AI and new technologies will bring to the labor market is still relatively small, but real. To prepare for continued adoption and advancements in the technologies, an immediate next step will involve the development of accurate, comprehensive and robust predictive models, using Gaussian Processes or long short-term memory (LSTM) based artificial recurrent neural networks (RNN), so as to provide guidance to workers, employers, and new graduates on skills and tasks of the future.

\bibliographystyle{aaai}
\bibliography{references.bib}

\begin{thebibliography}{}

\bibitem[\protect\citeauthoryear{Acemoglu and Restrepo}{2018}]{Acemoglu2018c}
Acemoglu, D., and Restrepo, P.
\newblock 2018.
\newblock {The race between man and machine: Implications of technology for
  growth, factor shares, and employment}.
\newblock {\em American Economic Review} 108(6):1488--1542.

\bibitem[\protect\citeauthoryear{Acemoglu and Restrepo}{2019}]{Acemoglu2019}
Acemoglu, D., and Restrepo, P.
\newblock 2019.
\newblock {Automation and New Tasks: How Technology Displaces and Reinstates
  Labor}.
\newblock {\em Journal of Economic Perspectives} 33(2):3--30.

\bibitem[\protect\citeauthoryear{Atalay \bgroup et al\mbox.\egroup
  }{2019}]{atalay2019evolution}
Atalay, E.; Phongthiengtham, P.; Sotelo, S.; and Tannenbaum, D.
\newblock 2019.
\newblock The evolution of work in the united states.
\newblock {\em American Economic Journal: Applied Economics}.

\bibitem[\protect\citeauthoryear{Autor and Dorn}{2013}]{AutorDorn2013}
Autor, D.~H., and Dorn, D.
\newblock 2013.
\newblock {The growth of low-skill service jobs and the polarization of the US
  Labor Market}.
\newblock {\em American Economic Review} 103(5):1553--1597.

\bibitem[\protect\citeauthoryear{Bartel, Ichniowski, and
  Shaw}{2007}]{Bartel2007}
Bartel, A.; Ichniowski, C.; and Shaw, K.
\newblock 2007.
\newblock {How Does Information Technology Affect Productivity? Plant-Level
  Comparisons of Product Innovation, Process Improvement, and Worker Skills}.
\newblock {\em Quarterly Journal of Economics} 122(4):1721--1758.

\bibitem[\protect\citeauthoryear{Beaudry, Green, and
  Sand}{2015}]{BeaudryGreenSand2015}
Beaudry, P.; Green, D.~A.; and Sand, B.~M.
\newblock 2015.
\newblock {The Great Reversal in the Demand for Skill and Cognitive Tasks}.
\newblock {\em Journal of Labor Economics} 34(S1):S199--S247.

\bibitem[\protect\citeauthoryear{Bishop}{2006}]{bishop2006pattern}
Bishop, C.~M.
\newblock 2006.
\newblock {\em Pattern recognition and machine learning}.
\newblock Springer.

\bibitem[\protect\citeauthoryear{Brynjolfsson and
  McAfee}{2014}]{BrynjolfssonMcAfee2014}
Brynjolfsson, E., and McAfee, A.
\newblock 2014.
\newblock {\em {The second machine age: Work, progress, and prosperity in a
  time of brilliant technologies}}.
\newblock WW Norton {\&} Company.

\bibitem[\protect\citeauthoryear{Brynjolfsson and
  McElheran}{2016}]{Brynjolfsson2016}
Brynjolfsson, E., and McElheran, K.
\newblock 2016.
\newblock {The Rapid Adoption of Data-Driven Decision-Making}.
\newblock {\em American Economic Review} 106(5):133--139.

\bibitem[\protect\citeauthoryear{Brynjolfsson, Mitchell, and
  Rock}{2018}]{Brynjolfsson2018b}
Brynjolfsson, E.; Mitchell, T.; and Rock, D.
\newblock 2018.
\newblock {What Can Machines Learn and What Does It Mean for Occupations and
  the Economy?}
\newblock {\em AEA Papers and Proceedings} 108:43--47.

\bibitem[\protect\citeauthoryear{de Troya \bgroup et al\mbox.\egroup
  }{2018}]{depredicting}
de~Troya, I.~M.; Chen, R.; Moraes, L.~O.; Bajaj, P.; Kupersmith, J.; Ghani, R.;
  Br{\'a}s, N.~B.; and Zejnilovic, L.
\newblock 2018.
\newblock Predicting, explaining, and understanding risk of long-term
  unemployment.
\newblock {\em NeurIPS Workshop on AI for Social Good}.

\bibitem[\protect\citeauthoryear{Deming and Kahn}{2018}]{Deming2018}
Deming, D., and Kahn, L.~B.
\newblock 2018.
\newblock {Skill Requirements across Firms and Labor Markets: Evidence from Job
  Postings for Professionals}.
\newblock {\em Journal of Labor Economics} 36(S1):S337--S369.

\bibitem[\protect\citeauthoryear{Deming and Noray}{2018}]{Deming2018a}
Deming, D.~J., and Noray, K.
\newblock 2018.
\newblock {STEM Careers and Technological Change}.
\newblock {\em NBER Working Paper}.

\bibitem[\protect\citeauthoryear{Duckworth, Graham, and
  Osborne}{2019}]{duckworth2019inferring}
Duckworth, P.; Graham, L.; and Osborne, M.
\newblock 2019.
\newblock Inferring work task automatability from {AI} expert evidence.
\newblock In {\em Proceedings of the AAAI/ACM Conference on AI, Ethics, and
  Society},  485--491.

\bibitem[\protect\citeauthoryear{Fleming \bgroup et al\mbox.\egroup
  }{2019}]{fleming2019future}
Fleming, M.; Clarke, W.; Das, S.; Phongthiengtham, P.; and Reddy, P.
\newblock 2019.
\newblock The future of work: How new technologies are transforming tasks.
\newblock {\em MIT-IBM Watson AI Lab}.

\bibitem[\protect\citeauthoryear{Gathmann and Schoenberg}{2010}]{Gathmann2010}
Gathmann, C., and Schoenberg, U.
\newblock 2010.
\newblock {How General Is Human Capital? A Task-Based Approach}.
\newblock {\em Journal of Labor Economics} 28(1):1--49.

\bibitem[\protect\citeauthoryear{Goldin and Katz}{2009}]{Goldin2007}
Goldin, C., and Katz, L.~F.
\newblock 2009.
\newblock {The Race between Education and Technology: The Evolution of U.S.
  Educational Wage Differentials, 1890 to 2005}.

\bibitem[\protect\citeauthoryear{Goos, Manning, and
  Salomons}{2014}]{GoosManningSalomons2014}
Goos, M.; Manning, A.; and Salomons, A.
\newblock 2014.
\newblock {Explaining Job Polarization: Routine-Biased Technological Change and
  Offshoring}.
\newblock {\em American Economic Review} 104(8):2509--2526.

\bibitem[\protect\citeauthoryear{Grace \bgroup et al\mbox.\egroup
  }{2018}]{grace2018will}
Grace, K.; Salvatier, J.; Dafoe, A.; Zhang, B.; and Evans, O.
\newblock 2018.
\newblock When will {AI} exceed human performance? {E}vidence from {AI}
  experts.
\newblock {\em Journal of Artificial Intelligence Research} 62:729--754.

\bibitem[\protect\citeauthoryear{Hemamou \bgroup et al\mbox.\egroup
  }{2019}]{hemamou2019hirenet}
Hemamou, L.; Felhi, G.; Vandenbussche, V.; Martin, J.-C.; and Clavel, C.
\newblock 2019.
\newblock Hire{N}et: a hierarchical attention model for the automatic analysis
  of asynchronous video job interviews.
\newblock {\em AAAI Conference on Artificial Intelligence (AAAI)}.

\bibitem[\protect\citeauthoryear{Hershbein and Kahn}{2018}]{Hershbein2018}
Hershbein, B., and Kahn, L.~B.
\newblock 2018.
\newblock {Do Recessions Accelerate Routine-Biased Technological Change?
  Evidence from Vacancy Postings}.
\newblock {\em American Economic Review} 108(7):1737--1772.

\bibitem[\protect\citeauthoryear{Makridakis, Wheelwright, and
  Hyndman}{2008}]{makridakis2008forecasting}
Makridakis, S.; Wheelwright, S.~C.; and Hyndman, R.~J.
\newblock 2008.
\newblock {\em Forecasting methods and applications}.
\newblock John wiley \& Sons.

\bibitem[\protect\citeauthoryear{Spitz-Oener}{2006}]{Spitz-Oener2006}
Spitz-Oener, A.
\newblock 2006.
\newblock {Technical Change, Job Tasks, and Rising Educational Demands: Looking
  outside the Wage Structure}.
\newblock {\em Journal of Labor Economics} 24(2):235--270.

\bibitem[\protect\citeauthoryear{Wajcman}{2017}]{wajcman2017automation}
Wajcman, J.
\newblock 2017.
\newblock Automation: {I}s it really different this time?
\newblock {\em The British journal of sociology} 68(1):119--127.

\end{thebibliography}

\clearpage


\appendix
\section{Appendix}

\vskip20pt
\subsection{Supplementary Section of Tables}

\begin{table}[h!]
\centering
\scriptsize
\begin{tabular}{l | r}
                        Task &  No. of Occupations \\
\hline
        Communication Skills &                 460 \\
           Computer Literacy &                 393 \\
       Organizational Skills &                 380 \\
                     Writing &                 372 \\
    Teamwork / Collaboration &                 364 \\
                  Scheduling &                 361 \\
             Detail-Oriented &                 342 \\
          Physical Abilities &                 338 \\
            Customer Service &                 336 \\
                     English &                 332 \\
                    Research &                 326 \\
             Problem Solving &                 323 \\
             Microsoft Excel &                 314 \\
       Written Communication &                 306 \\
                    Planning &                 304 \\
\hline
\end{tabular}
\caption{Tasks that appear in more than 300 occupations.
\label{table:task2soc}}
\end{table}

\begin{table}[h!]
\centering
\scriptsize
\begin{tabular}{l | r | r}
     SOC &  Occupation & No. of  \\
     & & Tasks\\
\hline \hline 
 15-1132 &      Software Developers, Applications   &  2312 \\  [1ex]
 11-9199 &      Managers, All Other   &  2113 \\ [1ex]
 41-4012 &      Sales Representatives,  &  \\
         &       Wholesale \& Manufacturing & 1310 \\ [1ex]
 15-1121 &      Computer Systems Analysts   &  1144 \\ [1ex]
 13-1111 &      Management Analysts   &  1144 \\ [1ex]
 11-9111 &      Medical \& Health Services Managers   &  1094 \\ [1ex]
 11-2021 &      Marketing Managers   &  1089 \\ [1ex]
 11-1021 &      General \& Operations Managers   &  1067 \\ [1ex]
 11-2022 &      Sales Managers   &  1065 \\
\hline
\end{tabular}
\caption{Occupations with more than 1000 unique tasks.
\label{table:soc2task}}
\end{table}

\begin{table}[h!]
\centering
\scriptsize
\begin{tabular}{l | r | r | r}
                         Task Clusters within IT &   High Wage &    Mid Wage &    Low Wage \\
\hline
\hline
       Artificial Intelligence &   0.0003118 &             &              \\
                      Big Data &   0.0007821 &             &              \\
           Scripting Languages &   0.0001187 &   -4.18e-05 &              \\
                     C and C++ &  -0.0001528 &  -0.0001776 &              \\
                     Scripting &   -0.000172 &    0.000148 &              \\
 SQL Databases and Programming &  -0.0001388 &  -0.0002113 &  -0.0005945 \\
         JavaScript and jQuery &     6.8e-05 &  -0.0004148 &  -0.0008842 \\
                          Java &  -0.0001515 &  -0.0001293 &    3.95e-05 \\
                 Cybersecurity &    4.75e-05 &   0.0004113 &  -0.0003633 \\
          Information Security &    1.06e-05 &    9.15e-05 &  -0.0002753 \\
               Cloud Solutions &   0.0002228 &    -3.8e-05 &  -0.0001886 \\
               Data Management &    7.12e-05 &   -1.89e-05 &  -0.0002002 \\
\hline
\end{tabular}
\caption{Normalized regression coefficients of task-shares of selected IT task clusters for HML Wage Occupations. 
\label{table:IT_HML_coeff} }
\end{table}


\begin{table}[h!]
\centering
\scriptsize
\begin{tabular}{l | r | r}
                                 Occupation Family & Health Care & Information \\
                                                   &             & Technology \\
\hline \hline
                            Management          &     4.7e-06 &              -1.17e-05 \\
          Community and Social Service          &   -3.26e-05 &              -1.31e-05 \\
 Healthcare Practitioners and Technical         &      -3e-06 &              -4.94e-05 \\
                    Healthcare Support          &    2.01e-05 &               4.73e-05 \\
             Personal Care and Service          &   0.0003122 &              0.0003667 \\
     Office and Administrative Support          &       2e-06 &              -5.99e-05 \\
     Business and Financial Operations          &    -4.8e-05 &              -5.57e-05 \\
    Life, Physical, and Social Science          &   -7.67e-05 &              -4.24e-05 \\
      Education, Training, and Library          &   -7.57e-05 &              -7.99e-05 \\
                    Protective Service          &   -9.62e-05 &              -6.47e-05 \\
  Food Preparation and Serving Related          &      -1e-05 &             -0.0002106 \\
 Building and Grounds Cleaning and Maintenance  &   -1.16e-05 &                4.6e-05 \\
                     Sales and Related          &  -0.0001142 &             -0.0001034 \\
    Transportation and Material Moving          &    -3.5e-06 &               1.06e-05 \\
             Computer and Mathematical          &     6.7e-06 &              -8.32e-05 \\
          Architecture and Engineering          &     1.8e-06 &              -5.59e-05 \\
                                 Legal          &   0.0001034 &               6.77e-05 \\
 Arts, Design, Entertainment, Sports, and Media &    0.000116 &              -7.65e-05 \\
           Construction and Extraction          &   0.0002009 &              0.0001889 \\
 Installation, Maintenance, and Repair          &    5.08e-05 &              -4.54e-05 \\
                            Production          &    4.23e-05 &              -4.88e-05 \\
\hline
\end{tabular}
\caption{Normalized regression coefficients of task-shares of Healthcare \& Information Technology task cluster families. 
\label{table:SCF_healthcare_IT_coeff} }
\end{table}

\begin{table}[h!]
\centering
\scriptsize
\begin{tabular}{l || r | r | r}
                                 Task Cluster Family & High Wage & Mid Wage & Low Wage \\
\hline\hline
                 Customer and Client Support &      0.98 &     0.72 &     1.98 \\
                          Industry Knowledge &      1.28 &     1.94 &     2.53 \\
                                       Sales &      1.08 &     2.02 &     1.21 \\
                                 Health Care &      0.73 &     0.66 &     2.46 \\
                  Supply Chain and Logistics &      0.45 &     1.12 &     1.63 \\
                              Administration &      0.65 &     0.58 &     1.21 \\
                                    Business &      0.46 &     0.77 &     2.40 \\
                      Education and Training &      1.11 &     1.65 &     1.70 \\
                                     Finance &      0.44 &     0.60 &     4.48 \\
                      Information Technology &      0.44 &     0.72 &     1.34 \\
                  Personal Care and Services &      1.73 &     2.24 &     1.53 \\
                             Human Resources &      0.61 &     1.50 &     2.22 \\
         Public Safety and National Security &      2.01 &     2.21 &     4.12 \\
              Marketing and Public Relations &      0.91 &     1.02 &     2.55 \\
                           Media and Writing &      0.44 &     1.23 &     4.32 \\
                Manufacturing and Production &      0.61 &     0.72 &     1.65 \\
               Architecture and Construction &      0.77 &     0.95 &     1.93 \\
                                       Legal &      0.85 &     1.79 &     5.89 \\
       Maintenance, Repair, and Installation &      0.68 &     0.66 &     1.67 \\
                                      Design &      0.91 &     1.72 &     7.82 \\
       Economics, Policy, and Social Studies &      1.24 &     3.14 &    16.00 \\
                                    Analysis &      0.96 &     1.23 &     4.81 \\
                        Science and Research &      0.98 &     1.23 &     9.11 \\
                                 Environment &      1.01 &     3.06 &     5.64 \\
                                 Engineering &      0.45 &     1.56 &     5.19 \\
                        Energy and Utilities &      1.87 &     2.24 &      \\
 Agriculture, Horticulture, \& Outdoors &      3.57 &     2.00 &     1.97 \\
                                    Religion &     11.38 &    10.81 &      \\
\hline
\end{tabular}
\caption{ Mean absolute percentage error for one-step ahead predictions of task-shares. 
\label{table:TS_HML_pred} }
\end{table}

\begin{table}[h!]
\centering
\tiny
\begin{tabular}{l | r | r | r}
                         Task Cluster Family &   High Wage &    Mid Wage &    Low Wage \\
\hline \hline
                              Administration &    2.25e-05 &   -3.67e-05 &    5.77e-05 \\
                                       Sales &   -2.75e-05 &    -6.7e-06 &   -5.03e-05 \\
                                 Environment &   -9.31e-05 &   -0.000181 &  -0.0001135 \\
                          Industry Knowledge &    1.26e-05 &     1.3e-05 &    1.23e-05 \\
                                      Design &    -7.6e-06 &  -0.0001254 &  -0.0003482 \\
                                    Religion &  -0.0004568 &  -0.0002736 &          \\
       Maintenance, Repair, and Installation &    8.87e-05 &    4.53e-05 &   0.0001067 \\
                                 Health Care &    -5.5e-06 &    2.87e-05 &     5.9e-05 \\
              Marketing and Public Relations &    1.47e-05 &   -6.59e-05 &    2.56e-05 \\
                                     Finance &   -2.63e-05 &   -6.32e-05 &   -1.91e-05 \\
         Public Safety and National Security &   -7.26e-05 &     3.8e-05 &    7.83e-05 \\
                Manufacturing and Production &   -2.73e-05 &     3.3e-06 &    3.59e-05 \\
                        Energy and Utilities &  -0.0002393 &    2.47e-05 &          \\
                      Information Technology &   -6.18e-05 &   -5.71e-05 &    -8.6e-05 \\
                  Personal Care and Services &    1.95e-05 &     8.1e-06 &    3.61e-05 \\
       Economics, Policy, and Social Studies &   -4.37e-05 &   -4.44e-05 &  -0.0002623 \\
                  Supply Chain and Logistics &    1.77e-05 &   -4.59e-05 &    3.16e-05 \\
                        Science and Research &   -6.52e-05 &   -7.92e-05 &  -0.0002512 \\
                                 Engineering &   -4.46e-05 &   -2.59e-05 &    8.83e-05 \\
                      Education and Training &    -7.8e-05 &    1.05e-05 &    -4.9e-05 \\
               Architecture and Construction &     8.3e-05 &    4.32e-05 &   -7.35e-05 \\
 Agriculture, Horticulture, and the Outdoors &    1.69e-05 &    7.94e-05 &    4.29e-05 \\
                             Human Resources &    7.56e-05 &    6.12e-05 &    0.000166 \\
                                       Legal &   -4.06e-05 &  -0.0001084 &  -0.0001222 \\
                           Media and Writing &   -2.85e-05 &   -8.86e-05 &  -0.0001016 \\
                                    Analysis &   -1.39e-05 &       6e-07 &    5.16e-05 \\
                 Customer and Client Support &    5.52e-05 &       8e-06 &     1.1e-05 \\
                                    Business &    -7.5e-06 &   -5.04e-05 &   0.0001131 \\
\hline
\end{tabular}
\caption{Normalized regression coefficients of task-shares of task cluster families across HML Wage Occupations. 
\label{table:SCF_HML_coeff} }
\end{table}

\end{document}